# Low temperature physical properties of Co-35Ni-20Mo-10Cr alloy MP35N®

*J. Lu\*, V. J. Toplosky, R. E. Goddard and K. Han*

*National High Magnetic Field Laboratory, Florida State University, 1800 E. Paul Dirac Dr. Tallahassee, FL 32310*

\* Corresponding author: junlu@magnet.fsu.edu, phone: 001-850-644-1678

**Abstract**

Multiphase Co-35Ni-20Mo-10Cr alloy MP35N® is a high strength alloy with excellent corrosion resistance. Its applications span chemical, medical, and food processing industries. Thanks to its high modulus and high strength, it found applications in reinforcement of ultra-high field pulsed magnets. Recently, it has also been considered for reinforcement in superconducting wires used in ultra-high field superconducting magnets. For these applications, accurate measurement of its physical properties at cryogenic temperatures is very important. In this paper, physical properties including electrical resistivity, specific heat, thermal conductivity, and magnetization of as-received and aged samples are measured from 2 to 300 K. The electrical resistivity of the aged sample is slightly higher than the as-received sample, both showing a weak linear temperature dependence in the entire range of 2 - 300 K. The measured specific heat $C_p$ of 0.43 J/g-K at 295 K agrees with a theoretical prediction, but is significantly smaller than the values in the literature. The thermal conductivity between 2 and 300 K is in good agreement with the literature which is only available above 77 K. Magnetic property of MP35N® changes significantly with aging. The as-received sample exhibits Curie paramagnetism with a Curie constant $C = 0.175$ K. While the aged sample contains small amounts of a ferromagnetic phase even at room temperature. The measured MP35N® properties will be useful for the engineering design of pulsed magnets and superconducting magnets using MP35N® as reinforcement.

## 1. INTRODUCTION

MP35N® is a nonmagnetic multiphase Co-35Ni-20Mo-10Cr alloy that has very high yield strength and excellent corrosion resistance [1]. In addition to its application in the chemical industries, its biocompatibility makes it favorable in many biomedical applications. In this paper, attention is drawn to a number of unique applications at cryogenic temperatures for high field magnets. MP35N® has been used in ultrahigh field pulsed magnet as a non-magnetic reinforcement material thanks to its high strength and high elastic modulus and cryogenic temperature compatibility [2]. It was also suggested to be a suitable material as a reinforcement or substrate of high $T_c$ superconductors [3]. A recently US patent by A. Otto and K. Yamazaki [4] uses MP35N® as reinforcement laminations in high $T_c$ $Bi_2Sr_2Ca_2Cu_3O_{10+x}$ superconducting tapes (Bi-2223). Such reinforced Bi-2223 is much stronger than previously possible, therefore a significant improvement in field strength of magnets made by Bi-2223 is expected. For these applications in pulsed and superconducting magnets, the thermal, electrical and magnetic properties of MP35N® at cryogenic temperatures are very important to magnet engineering design such as superconducting magnet quench analysis, as well as magnet manufacturing and operation. Although some typical room temperature properties are available as listed in Table I, low temperature physical properties have not been reported. Therefore in this paper, we present our measurements of its specific heat capacity, thermal conductivity, electrical resistivity and magnetization from 2 K to 300 K.

One additional interest is the mechanism of age hardening in MP35N®. MP35N® is a work hardening material with a secondary hardening by aging at around 600 °C which results in an additional significant increase in strength [2]. Much research has been focused on understanding the aging mechanism and optimizing the aging time and temperature. The microstructures of aged samples, in particular, have been studies since the 1990s [2], [6] - [14]. However, the existence of precipitations is still debatable due to the limited accuracy in high spatial resolution chemical analysis. It is hoped that accurate physical property measurements may shed a light on the mechanism of the age-hardening which is of interest for fundamental material science.

## 2. EXPERIMENTAL

2.1 The material

MP35N® used for all measurements (except for thermal conductivity measurements) was casted by Carpenter Technology, and rolled to 0.145 mm (nominally 0.127 mm) thick sheets by H. C. Starck, Euclid, OH (Lot# 25123) with a total cold-work of 65%. For the thermal conductivity measurement, 0.145 mm thick foil sample resulted in significant measurement error especially at temperatures above 200 K primarily due to radiation heat loss that is difficult to correct [15]. Therefore a 6.86 mm diameter MP35N® rod was purchased from Tech Steel and Materials Inc. The cold-work of this rod was estimated to be 65-74% based on our hardness measurements.

The chemical composition of MP35N® sheet is measured by X-ray fluorescence (XRF) spectrometry by Carpenter Technology. The composition of the MP35N® rod used for thermal conductivity measurement is measured by Energy Dispersive X-ray Spectroscopy (EDS) in a scanning electron microscope. Table II compares the compositions of these samples with the nominal composition of MP35N®. The as-received MP35N® sheet is aged at 550 °C for 8 hours in an argon environment and then quickly air-cooled to room temperature.

2.2 Mechanical tensile test

All tensile testing was performed on 'dog-bone' shaped samples as per ASTM E8 M. Room temperature tests were performed on an MTS Criterion screw-driven model with a 1" MTS clip-on extensometer and 5 kN capacity load cell. All 77 K tests were performed on a 250 kN servo-hydraulic test machine with a 1" clip-on ETS extensometer and 250 kN load cell. For both tests, the actuator stroke rate was 0.5 mm/min. The specimen was loaded to 1.5% strain, unloaded and then reloaded to failure. Modulus of elasticity, yield strength using the 0.2% offset method, ultimate strength and elongation to failure were recorded and then analysed.

2.3 Resistivity measurement

The electrical resistivity versus temperature and magnetic field measurements were performed on a Quantum Design's physical property measurement system (PPMS) operating in a temperature range of 2 – 300 K and field range of 0 – 9 T. Samples are typically strips of 0.145 mm thick, 2 mm wide and 12 mm long. Sample length direction coincides with the sheet rolling direction. Four 0.1 mm diameter pure silver wires were attached to each sample with silver paint for a four-probe measurement. The measurements were made with a 7.5 Hz alternating positive and negative excitation current of 1 mA. The system is calibrated with an internal calibration resistor.

Since the sample length in the PPMS is typically limited to 15 mm, the relative measurement accuracy is limited by the uncertainty in sample dimension measurements. To reduce the uncertainty, samples of larger size (200 x 12.7 x 0.125 mm$^3$) are measured at room temperature (295 K) with a distance of 100 mm between voltage taps. This reduces measurement uncertainty to about 1%. A correction factor is applied to the resistivity measured by the PPMS, so that at 295 K and zero magnetic field the resistivity is in agreement with that measured by the large sample.

2.4 Specific heat capacity ($C_p$) measurement

$C_p$ is measured by the PPMS heat capacity option which was calibrated with a gold sample provided by Quantum Design in a temperature range of 2 – 300 K. Before the measurement of each sample, an addenda file for the sample holder was measured. An as-received MP35N® sheet was cut into a sample of 4.1 mg and mounted to the holder. The measurement is in a vacuum better than 9 x 10$^{-6}$ Torr and in a temperature range of 2 – 300 K. For each $C_p$ measurement, a heat pulse is applied to the sample, and the temperature response of the sample as a function of time is fitted by a model with $C_p$ as a fitting parameter. Details on PPMS heat capacity measurement and the model for fitting the temperature response curve can be found in [16]. The typical measurement uncertainty is estimated to be about +/-5%.

2.5 Thermal conductivity measurement

The thermal conductivity is measured by the PPMS thermal transport option which is calibrated by a 201 Ni alloy sample provided by Quantum Design. Four gold-plated copper wires are attached to an as-received MP35N® sample of 1 x 4 x 8 mm$^3$ by silver epoxy which is cured at 150 °C for 3 minutes. These four wires were connected to a semiconductor heater, two Cernox temperature sensors, and a system temperature terminal respectively. Thermal conductivity was measured in continuous mode. In this mode, the system applies a heat pulse, and thermal conductivity is obtained by fitting the temperature response measured by both temperature sensors while the background temperature was continuously ramped from 2 to 300 K slowly. During the measurement, the vacuum is better than 9 x 10$^{-6}$ Torr, and the magnetic field is zero. The measurement error is estimated to be ± 5% [17].

2.6 Magnetization measurements

Magnetization versus temperature and magnetic field is measured by the vibrating sample magnetometer (VSM) in the PPMS. The sensitivity of the VSM is in the order of 10$^{-6}$ emu [18]. For each measurement, a sample of 23.6 mg (about 4 x 4 x 0.145 mm$^3$) is glued to a quartz sample rod by GE 7031 varnish so the applied magnetic field was either parallel or perpendicular to the rolling direction. Both as-received and aged samples were measured. The vibration frequency was 40 Hz and the amplitude was 2 mm. Measurement was performed between 2 and 300 K and up to 9 T magnetic field.

## 3. RESULTS AND DISCUSSIONS

### 3.1. Mechanical properties

The mechanical properties were tested at 295 K and 77 K for both as-rolled and aged samples. The modulus of elasticity $E$, 0.2% yield strength (YS), ultimate tensile strength (UTS), and elongation at failure ($e_f$) are listed in Table III, where AR, AG, LD and TD denote as-received, aged, longitudinal direction and transverse direction respectively. Each value in the table is an average from a group of 3 – 10 samples. The measurement uncertainty presented is the standard deviation within the group.

Aging increases strength in both 295 K and 77 K for both longitudinal and transverse directions. It increases yield strength somewhat more than ultimate tensile strength at both temperatures. At 295 K in longitudinal direction, for example, aging increases YS by 27% or 411 MPa, but UTS only by 23% or 368 MPa. Whereas the elongation is reduced by aging. As expected, both YS and UTS are higher at 77 K compared with those at 295 K. The effect of cryogenic temperature, however, is slightly less than the effect of aging. For an as-received sample, from 295 K to 77 K, UTS increases by about 15%, which is less than 23% increase induced by aging.

### 3.2. Electrical resistivity

Our room temperature measurements show no significant difference in resistivity between parallel and perpendicular to the sample's rolling direction within measurement accuracy. The resistivity versus temperature is shown in Fig. 1(a). The measured value is in reasonable agreement with another source [19] (also plotted in Fig. 1(a)), and comparable to that of stainless steels and other Ni alloys [20], [21]. The resistivity ratio between room temperature and liquid helium temperature is only 1.1, which is comparable to that of Constantan which is known for its low resistivity temperature coefficient [24]. As expected, from 300 K to 20 K the resistivity decreases nearly linearly with decreasing temperature. This is due to a decrease in phonon-electron scattering as described by the Bloch-Gruneisson formula [22]. However, below 20 K when phonon density becomes very low, it continuously decreases with decreasing temperature at the same rate. This phenomenon might be related to the interaction between electrons and magnetic dipoles from transition metal elements Cr, Ni, Co [22]. As shown in Fig. 1(a), the aged sample has higher resistivity, which coincide with the fact that aged material has higher mechanical strength [11]. This suggests that the aging process results in a slightly less homogeneous system that has higher strength and higher resistivity. The resistivity versus magnetic field was measured for both as-received and aged samples. As shown in Fig. 1(b) for an aged sample, the magnetoresistivity at 300 K and 77 K is negligible. At 4.2 K, the magnetoresistivity is $1.106 \pm 0.007 \times 10^{-9}$ $\Omega$-m/T and $7.85 \pm 0.05 \times 10^{-10}$ $\Omega$-m/T for as-received and aged sample respectively. These values are more than an order of magnitude higher than that of pure copper (with RRR = 100) which has a value of 4.1 x $10^{-11}$ Ω-m/T. [23].

### 3.3. Thermal expansion

Thermal expansion of MP35N® at low temperatures has been measured by a capacitive dilatometer [2]. We reproduced the results in [2] and plotted it in Fig. 2. Thermal expansion of Cu and 316 stainless steel [24], commonly used in superconducting magnets, are also plotted for comparison. Thermal expansion of MP35N® is significantly smaller than that of copper and 316 stainless steel, but is comparable with other Ni based super-alloys such as Hastelloy C-276 [20] and Haynes 242 [21].

### 3.4. Specific heat capacity $C_p$

Fig. 3 shows the measured $C_p$ versus temperature of an as-received sample in zero magnetic field. Data from other sources [25], [26] at room temperature and those of Hastelly C-276 [20] are also plotted for comparison. As expected, $C_p$ increases with $T$ in agreement with Debye's model [21]. No anomaly is observed.
`
It is noticed that our $C_p$ value 0.424 J/g-K at 295 K is considerably lower than 0.75 J/g-K tabulated in the US Military Handbook [25] which is also adopted in 2006 by the European Space Agency (ESA) in the European Space Materials database (ESMDB) and Medical Device Handbooks [26]. Some commercial sources provide room temperature $C_p$ value to be 0.502 J/g-K [27] which is still higher than our value. Since the original source of the above mentioned $C_p$ values are not given, it is difficult to investigate the cause of this considerable discrepancy. However, our value agrees with an estimate by Dulong-Petit law, which says for metallic materials at room temperature, $C_p = 3R$, where $R$ is the gas constant 8.31 J/K-mole. Based on the average molar weight of MP35N® calculated from the measured composition in Table II, the estimated value is $C_p = 0.423$ J/g-K, in a very good agreement with our measured $C_p$ value.

### 3.5. Thermal conductivity

The thermal conductivity versus temperature data for the as-drawn sample is plotted in Fig. 4. For comparison, MP35N® thermal conductivity data reproduced from [25] (available only at above 77 K) and thermal conductivity of Hastelloy C-276 [20] are also plotted. For MP35N®, the agreement between our data and data in [25] is very good. Compared with Hastelloy C-276, MP35N® has considerably lower thermal conductivity especially at temperatures below 77 K.

### 3.6. Magnetic properties

For an as-received sample, $\chi$ shows a typical Curie paramagnetic behavior, i.e., it increases monotonically with decreasing temperature (Fig. 5(a)). Detailed data analysis shows that if we assume a small $T$ independent component $\chi_c = 6.8 \times 10^{-4}$, which may be attributed to Pauli paramagnetism of conduction electrons, the rest of susceptibility $\chi' = \chi - \chi_c$ can be fitted very well by the Curie-Weiss law,

$$\chi' = C/(T-T_C) \qquad (1)$$

where $C$ and $T_C$ are the Curie constant and Curie temperature respectively. According to (1), $1/\chi'$ versus $T$ plot is a straight line above the Curie temperature. This lineality in most of the temperature range is confirmed as shown in the inset of Fig. 5(a). By fitting $1/\chi'$ versus $T$ data between 25 and 50 K with equation (1), we obtained the Curie constant $C = 0.175$ K and Curie temperature $T_C = 6.4$ K which is in agreement with graphically extrapolating the linear section of $1/\chi'$ vs. $T$ to $1/\chi' = 0$ as depicted in Fig. 5(b). This implies that at least part of this sample transitions to a weak ferromagnetic phase at Curie temperature $T_C = 6.4$ K.

When looking into ferromagnetism of a transition metal alloy like MP35N®, it is possible that small ferromagnetic clusters create superparamagnetism which is featured by a blocking behavior where $\chi$ of a field-cooled (FC) and a zero-field-cooled (ZFC) measurement are different below a characteristic temperature. In order to investigate this possibility, magnetization in a 1000 Oe field was measured with increasing temperature in both the FC and ZFC conditions. The $\chi$ vs. $T$ of FC and ZFC experiments are identical (not shown), which indicates the absence of superparamagnetism.

The magnetic susceptibility $\chi$ versus $T$ of the aged sample as shown in Fig. 5(c) is much weaker and shows significantly less $T$ dependence than the as-received sample (Fig. 5(a)). The $1/\chi$ versus $T$ curve, as shown in the inset of Fig. 5(c), is not a straight line indicating a complicated behavior considerably different from the Curie paramagnetism.

Magnetization versus field loops were also measured at 300, 77 and 4.2 K for both as-received (Fig. 6(a)) and aged samples (Fig. 6(b)). For the as-received sample at 77 and 300 K, *M* is approximately proportional to the magnetic field as it would be for a paramagnetic material. While at 4.2 K, as shown in the inset of Fig. 6(a), it shows hysteresis behavior indicating a ferromagnetic component with a coesivity of 100 Oe, a remnant magnetization of 1.6 kA/m, and a saturation magnetization of 12.6 kA/m. This is consistent with the $1/\chi$ versus *T* curve which indicates a weak ferromagnetic transition below 6.4 K. In contrast, the aged sample has much smaller total magnetization signal at all temperatures (Fig 6 (b)), but as shown in the inset, a small ferromagnetic contribution is clearly discernable with a coesivity of 200 Oe, a remnant magnetization of 0.12 kA/m at room temperature, and moderately increases with decreasing temperature.

Despite the existence of ferromagnetism in this material, the total magnetization remains very small, and should not be a concern for its application in high field magnets. For the aged sample, it is also noticed that the difference between the field parallel and perpendicular to the rolling direction is not significant for both as-received and aged samples.

Although the appearance of room temperature ferromagnetism after aging might provide a clue for the age hardening mechanism, at this point, there is no direct evidence to link the appearance of ferromagnetism with microstructure change after the aging heat treatment. Detailed microstructure analysis using transmission electron microscopy is underway to study these samples.

## 4. Conclusions

Electrical resistivity, specific heat capacity, thermal conductivity and magnetization of MP35N® are measured from 2 to 300 K. Resistivity of the aged sample is slightly higher than that of as-received sample, and neither is sensitive to temperature. The magnetoresistance is negligible at 77 and 300 K. The measured specific heat capacity at room temperature agrees very well with Dulong-Petit prediction, but is considerably smaller compared with those in the literature. There is significant difference in magnetization between as-received (65% cold-work) and aged samples. The as-received sample is paramagnetic above 6.4 K, but weakly ferromagnetic below 6.4 K. In contrast, the aged sample is slightly ferromagnetic at room temperature. The physical properties of both as-received and aged MP35N® is comparable to other Ni based alloys such as Hastelloy C-276, and suitable for applications in high field magnets at cryogenic temperatures.

## 5. Acknowledgement

The authors thank Dr. Antonius de Rooij, consultant of the European Space Agency on Materials Technology, for helpful discussions on heat capacity and thermal conductivity of MP35N®. Financial supports of the National Science Foundation under grant of DMR-0084173 and the State of Florida are gratefully acknowledged.

# 6. Reference

[1] G.S. Smith, US patent no. 3,356,542.

[2] K. Han, A. Ishmaku, Y. Xin, H. Garmestani, V.J. Toplosky, R.P. Walsh, C. Swenson C, B. Lesch, H. Ledbetter, S. Kim, M. Hundley, and J. R. Sims Jr., IEEE Trans. Appl. Supercond. 12 (2002)1244-47.

[3] K. Han, V.J. Toplosky, J. Lu, Y. Xin, R.E. Goddard, R.P. Walsh, and I.R. Dixon, IEEE Trans. Appl. Supercond. 23 (2013) 7800204.

[4] A. Otto, K. Yamazaki, US patent no. 20160141080 A1.

[5] http://cartech.ides.com/datasheet.aspx?i=102&e=3

[6] A. Graham, J. Youngblood, Metall. Mater. Trans. B 1 (1970) 423–30.

[7] R. Singh, R. Doherty, Metall. Mater. Trans. A 23 (1992) 307–19.

[8] S. Asgari. E. El-Danaf, E. Shaji, S.R. Kalidindi and R.D. Doherty, Acta Mater. 46 (1988) 5795–806.

[9] A. Ishmaku, K. Han, Mater. Charact. 47 (2001) 139–48.

[10] A. Ishmaku, K. Han, J. Mater. Sci. 2004;39:5417–20.

[11] V.J. Toplosky, K. Han, AIP Conference Proceedings 125 (2012) 1435.

[12] D. Sorensen, B.Q. Li, W.W. Gerberich, K.A. Mkhoyan, Acta Mater. 63 (2104) 63–72.

[13] M.J.N.V. Prasad, M.W. Reiterer, K.S. Kumar, Mater. Sci. Eng. 610 (2014) 326–337.

[14] M.J.N.V. Prasad, M.W. Reiterer, K.S. Kumar, Mater Sci Eng A 636 (2015) 340–351.

[15] Neil Dilley, private communications.

[16] PPMS Heat Capacity Option User's manual, 11[th] Ed. Quantum Design, 2004.

[17] PPMS Thermal Transport Option User's manual and application notes, 2nd Ed. Quantum Design, 2002.

[18] PPMS Vibrating Sample Magnetometer (VSM) Option User's manual, 5th Ed. Quantum Design, 2011.

[19] http://www.matweb.com/ H.C. Starck MP35N® Nickel/Cobalt/Chromium/Molybdenum Alloy.

[20] J. Lu, E.S. Choi, H.D. Zhou, J. Appl. Phys. 103 (2008) 064908.

[21] J. Lu, K. Han K, E.S. Choi, Y. Jo, L. Balicas, Y. Xin, J. Appl. Phys. 101 (2007) 123710.

[22] H. M. Rosenberg, *Low Temperature Solid State Physics,* Oxford University Press, London, 1963.

[23] N. J. Simon, E.S. Drexler, R.P. Reed, NIST Monograph 177, U.S. Government printing office, Washington 1992.

[24] J.E. Jensen, W.A. Tuttle, R.B. Steward, H. Brechna, A.G. Prodell, Brookhaven National Laboratory Selected Cryogenic Data Notebook, volume 1, 1980

[25] Metallic Materials and Elements for Aerospace Vehicle Structures, US Department of Defense, MIL-HDBK-5H, 1998.

[26] Materials and Coatings for Medical Devices Cadiovascular, ASM international, 2009.

[27] https://www.zapp.com/fileadmin/downloads/01-Produkte/Oel-und-Gas/Alloy-MP35N-e_USA_01-13.pdf

Table I Room temperature properties of MP35N® [5]

| Properties | Value |
| --- | --- |
| Density | 8.43 g/cc |
| Resistivity | $1.03 \times 10^{-6}$ Ω-m |
| CTE | $1.28 \times 10^{-5}$ K$^{-1}$ |
| Thermal conductivity | 11.2 W/m-K |
| Magnetic permeability | 1.00092 |
| Young's modulus | 235 GPa |
| Ultimate tensile strength (50 % Cold Rolled, aged at 550 °C) | 1830 MPa |

Table II The chemical composition of MP35N® in weight percent

|  | Co | Ni | Cr | Mo | Si | Fe | Ti | Mn | Al |
|---|---|---|---|---|---|---|---|---|---|
| Nominal [5] | balance | 33-37 | 19-21 | 9-10.5 | <0.15 | <1.00 | <1.00 | <0.15 |  |
| Sheet | 34.52 | 35.7 | 20.4 | 9.4 | 0.01 | 0.04 | 0.001 | 0.01 |  |
| Rod for Thermal conductivity | 31.6 | 36.2 | 20.8 | 8.6 | - | 1.04 | 0.82 | - | 0.90 |

Table III Mechanical properties

| | | 295 K | | | | 77 K | | | |
|---|---|---|---|---|---|---|---|---|---|
| | | E (GPa) | YS (MPa) | UTS (MPa) | $e_f$ (%) | E (GPa) | YS (MPa) | UTS (MPa) | $e_f$ (%) |
| AR | LD | 177 ±10 | 1514 ±10 | 1592 ±8 | 2.4 ±0.4 | 194 ±3 | 1914 ±50 | 2048 ±26 | 6.7 ±1.5 |
| AR | TD | 211 ±9 | 1481 ±26 | 1719 ±20 | 6.3 ±0.8 | 234 ±7 | 1862 ±43 | 2160 ±37 | 9.5 ±1.7 |
| AG | LD | 210 ±5 | 1925 ±15 | 1960 ±27 | 1.5 ±0.1 | 226 ±4 | 2237 ±38 | 2331 ±38 | 2.4 ±0.2 |
| AG | TD | 246 ±3 | 1990 ±37 | 2124 ±14 | 1.8 ±0.1 | 261 ±5 | 2334 ±59 | 2499 ±35 | 5.5 ±0.7 |

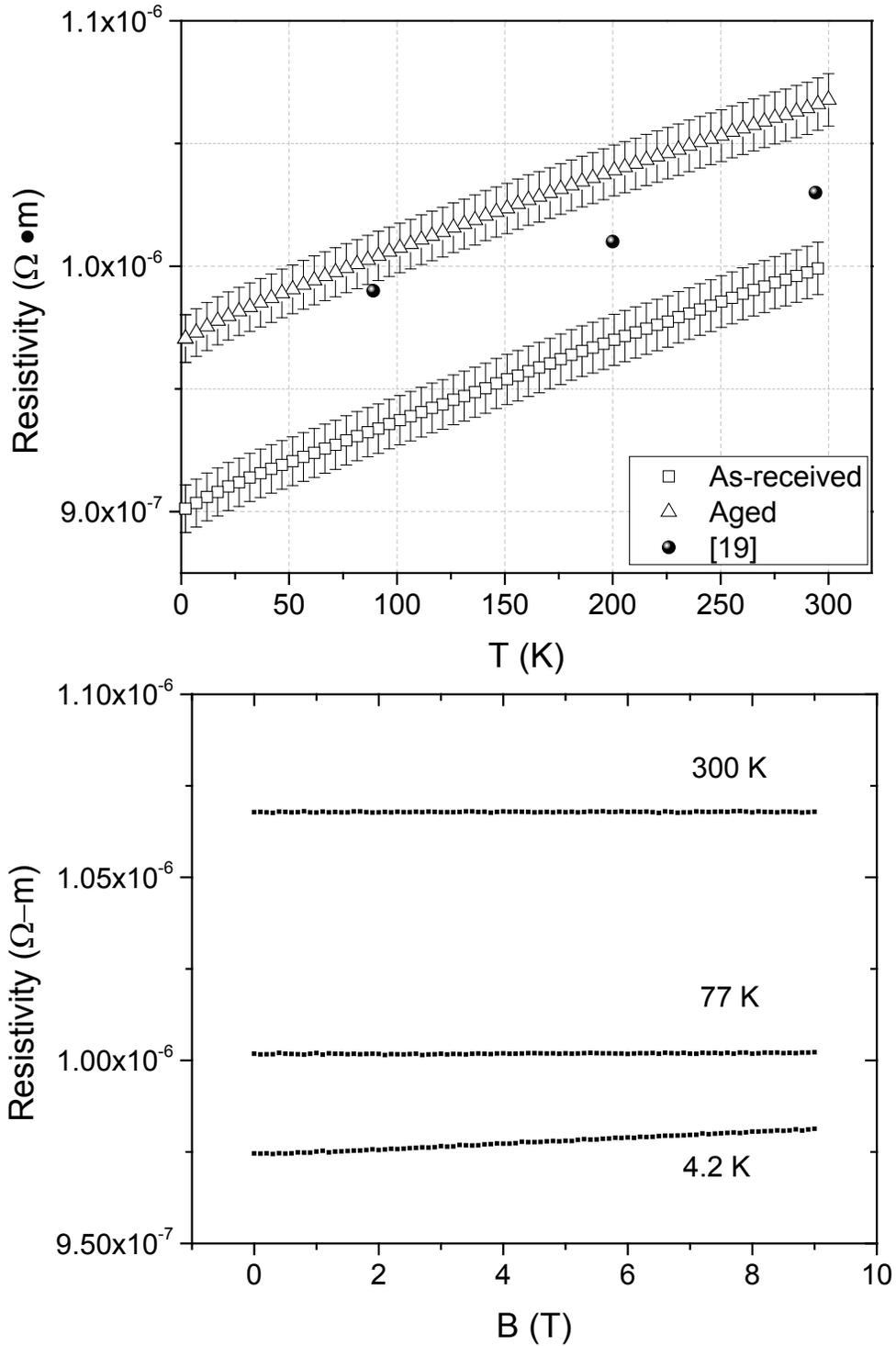

Fig. 1. (a) Electrical resistivity versus temperature for an as-received and an aged sample. Both samples are configured so that the electrical current flows along the rolling direction. The data from [19] is shown for comparison. (b) Magnetoresistivity of an aged sample.

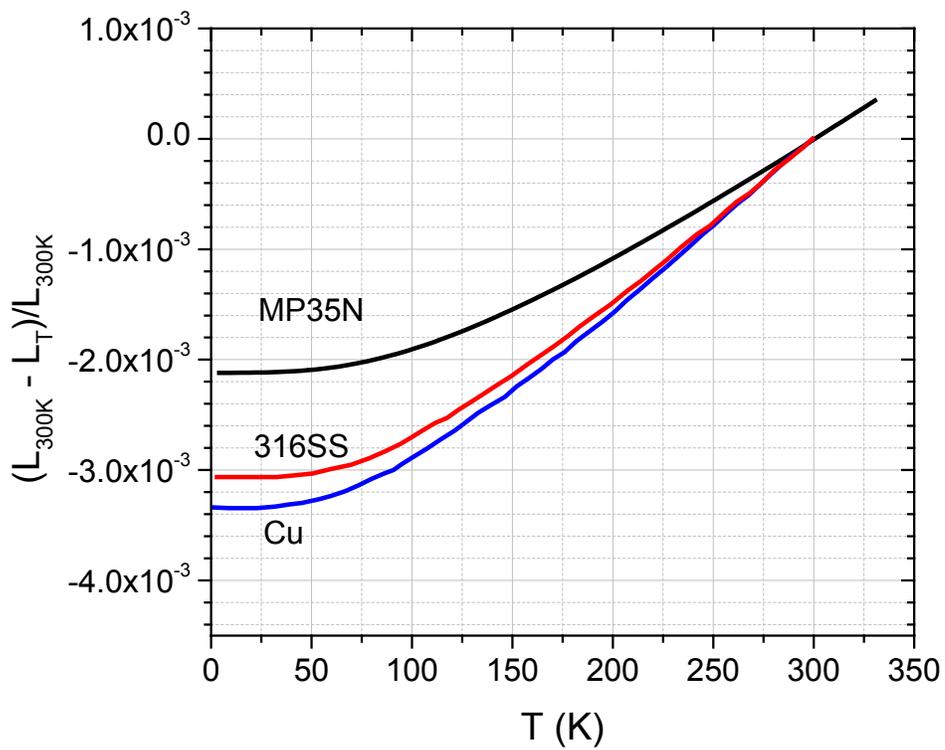

Fig. 2. Thermal expansion of MP35N® at low temperatures, integrated from data reproduced from [2]. 316 stainless steel and copper data from [24] are also plotted for comparison.

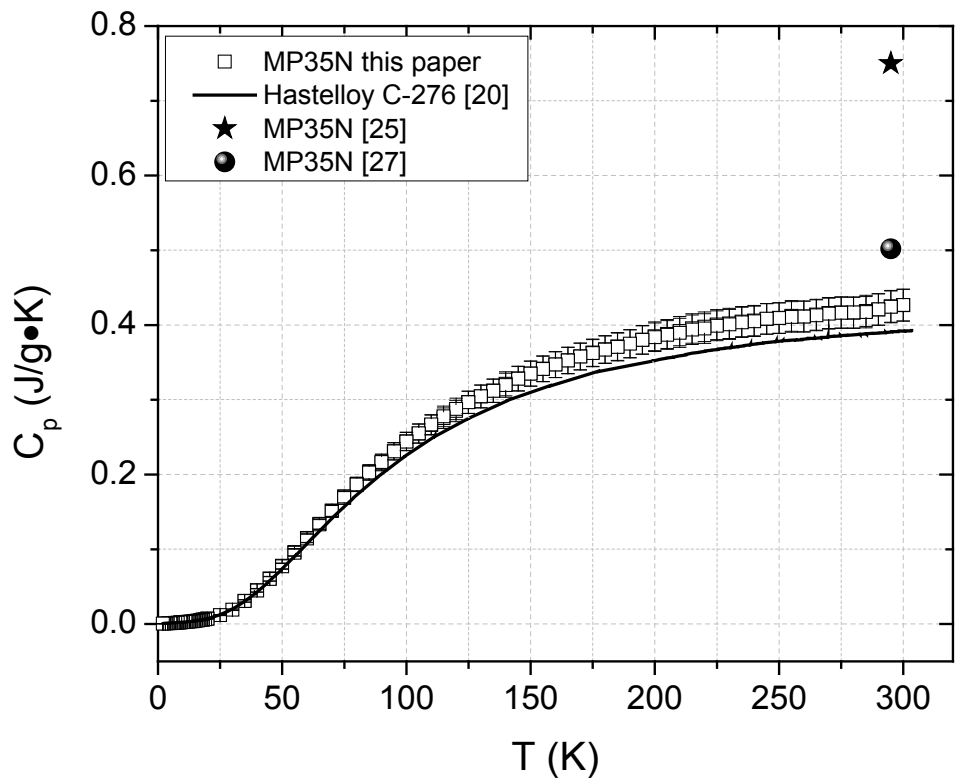

Fig. 3. Specific heat capacity of an as-received sample as a function of temperature. Room temperature data from [25] and [27] as well as data for Hastelloy C-276 [20] are also plotted for comparison.

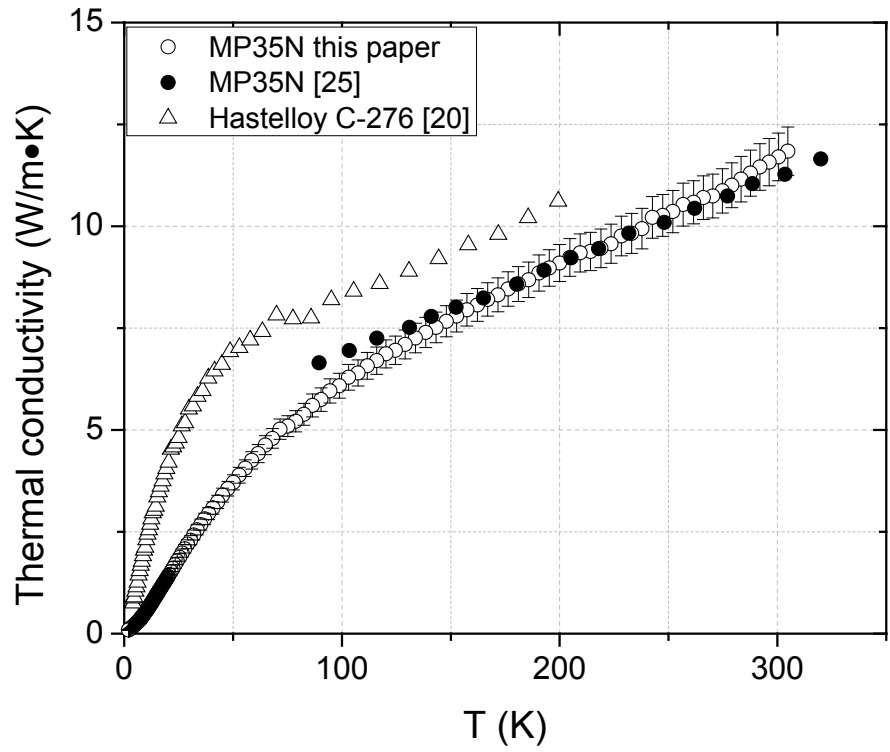

Fig. 4. Thermal conductivity of the cold-drawn sample versus temperature from 2 to 300 K. Data from [25] and Hastelloy C-276 data from [20] are also plotted for comparison.

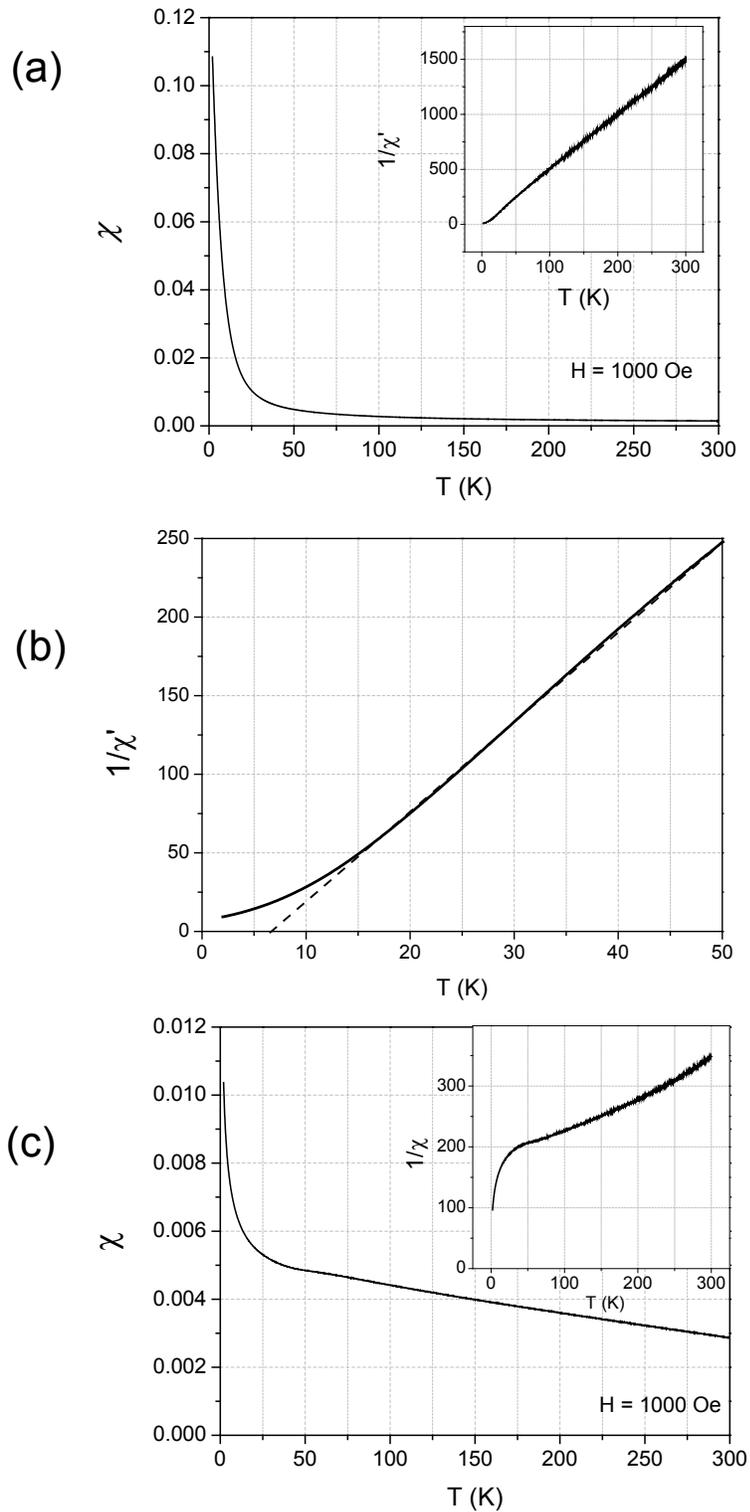

Fig .5. (a) Dimensionless magnetic susceptibility $\chi$ vs. $T$ for an as-received sample. Field parallel to rolling direction. When a small $T$ independent component is subtracted, $1/\chi'$ is linear with $T$ above 20 K as shown in the inset. (b) a close-up of $1/\chi'$ vs. $T$ from the inset of (a), where the extrapolation of the straight line intercepts the x-axis at $T = 6.4$ K. (c) $\chi$ vs. $T$ for the aged sample. The inset is $1/\chi$ vs. $T$ curve that cannot be fitted with a straight line.

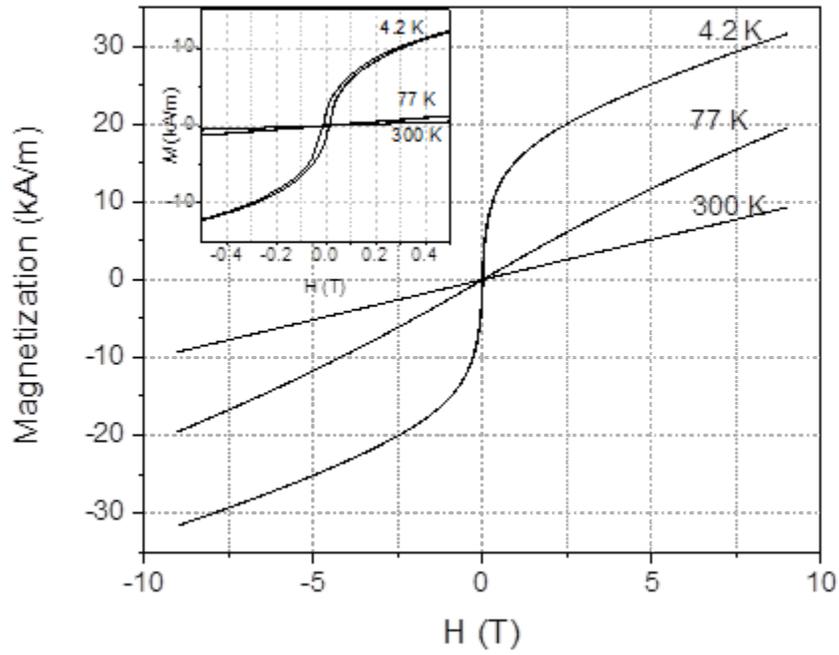

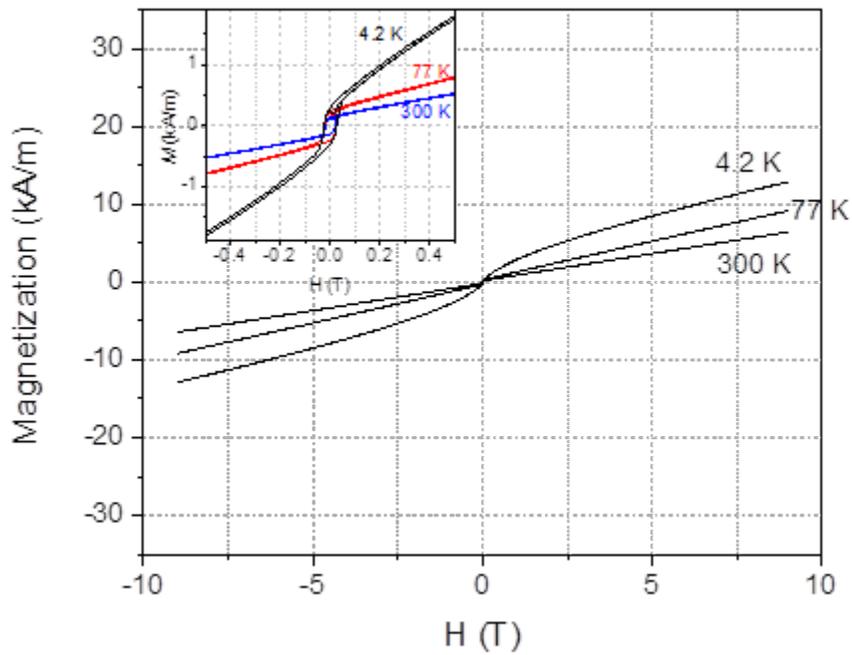

Fig. 6. Magnetization versus magnetic field for (a) an as rolled sample, and (b) an aged sample. Both with field parallel to rolling direction. The inset in each figure is a close-up showing hysteresis.

24